\documentclass[twocolumn,aps,amssymb,prl,showpacs]{revtex4}
\usepackage{graphicx}
\usepackage{amsmath}
\usepackage{float}

\newcommand{\RR}{\right}
\newcommand{\LL}{\left}
\newcommand{\m}{\mathrm}

 \bigskip
\begin{document}
\title{Accessing nanomechanical resonators via a fast microwave circuit}

\author{Mika A. Sillanp\"a\"a}
\email{Mika.Sillanpaa@iki.fi}
\author{Jayanta Sarkar}
\author{Jaakko Sulkko}
\author{Juha Muhonen}
\author{Pertti J. Hakonen}
\affiliation{Helsinki University of Technology Low Temperature
Laboratory Puumiehenkuja 2B, Espoo, FIN-02015 HUT Finland}

\begin{abstract}
The measurement of micron-sized mechanical resonators by
electrical techniques is difficult, because of the combination of
a high frequency and a small mechanical displacement which
together suppress the electromechanical coupling. The only
electromagnetic technique proven up to the range of several
hundred MHz requires the usage of heavy magnetic fields and
cryogenic conditions. Here we show how, without the need of either
of them, to fully electrically detect the vibrations of conductive
nanomechanical resonators up to the microwave regime. We use the
electrically actuated vibrations to modulate an $LC$ tank circuit
which blocks the stray capacitance, and detect the created
sideband voltage by a microwave analyzer. We show the novel
technique up to mechanical frequencies of 200 MHz. Finally, we
estimate how one could approach the quantum limit of mechanical
systems.
\end{abstract}

\maketitle


Micro- and nanomechanical systems \cite{clelandbook,RoukesReview}
are increasingly finding use in various sensor applications, where
the vibrations of clamped beams or cantilevers are affected by the
measured quantities. The detection of mechanical vibrations at the
submicron scale in such systems gets notoriously difficult. This
essentially stems from the fact that the transduction of
mechanical motion into the engineering world based typically on
\emph{electrical} measurement techniques is not easy. Also, the
detection of higher mechanical resonant frequencies $f_0$ becomes
increasingly difficult, since the physical size shrinks roughly
inversely with increasing frequency. On the other hand, the
smallest nanomechanical resonators are the most interesting ones,
as sensors due to their small active mass \cite{NRsensor}, or, for
the benefit of basic research, for observing quantum-mechanical
phenomena in the collective mechanical degree of freedom
\cite{Schwab03,cleland04}.

The backbone for electronic readout of nanomechanical resonators
(NR) has been the magnetomotive method
\cite{YurkeMagn94,cleland96Si} or its variants
\cite{RoukesBalance} which have been proven above 500 MHz
\cite{pashkinNEMS}. Here, the current-carrying beam is vibrating
in a sizable magnetic field, thus inducing an electromotive force
which can be read out by a network analyzer. However, the method
has a practical constraint limiting its general applicability,
since typically a 1-10 Tesla field and hence superconducting
magnets and 4 Kelvin operation are needed.

Apart from simple readout, the most significant motivation for the
investigation of sensitive and fast detection methods for NR is
the ongoing quest towards experimentally reaching their quantum
ground state \cite{lahaye04,Naik06,lehnert08}. The quantum limit
necessitates, first of all, a low temperature $T$ such that $h f_0
\gtrsim k_B T$. New materials and fabrication techniques recently
paved the way to breaking $f_0 = 1$ GHz \cite{RoukesGHz},
corresponding to $T \sim 50$ mK which is relatively easily
attainable using standard dilution refrigeration techniques. The
prerequisites for the detection method, however, become
formidable. The magnetomotive readout appears not suitable for
pursuing the quantum limit, since the high field suppresses
superconductivity, which also gives rise to Joule heating.
Similarly, a frequently applied piezoelectric actuation may become
problematic under these circumstances.

Already from the point of view of practical benefit of either the
research scientist or, in particular, for a multitude of
high-frequency sensor applications, a straightforward electronic
readout would be valuable, which would desirably work up to $f_0
\sim 1$ GHz, and without the use of cumbersome high Tesla magnets.
Truitt \emph{et al.} recently took a first step to this direction
\cite{SchwabRNEMS}. They took advantage of the fact that at the
drive frequency a driven NR looks like a series electrical $RLC$
resonator, for which they improved the impedance match to $Z_0 =
50 \, \Omega$ by coupling the NR to an electrical $LC$ matching
circuit resonant with the mechanical mode.

In our work, we demonstrate a fully electric readout protocol
which uses an external tank circuit in order to eliminate the
external wiring capacitances which otherwise would mask the tiny
capacitance modulation which comes from the motion of the metallic
NR. In the present work, the $LC$ frequency $f_{\m{LC}} \gg f_0$
can be chosen high enough to ensure bandwidths up to the range of
1 GHz, independent on $f_0$. Similar techniques where an external
resonant ("tank", meaning energy storage) circuit have been used
in order to enhance the detection bandwidth are well-known in the
mesoscopic electron transport community
\cite{rfset,CSET,sillanpaa04}. Estimates show that the method is
promising in approaching the quantum limit.

\begin{figure*}
\includegraphics[width=15.0cm]{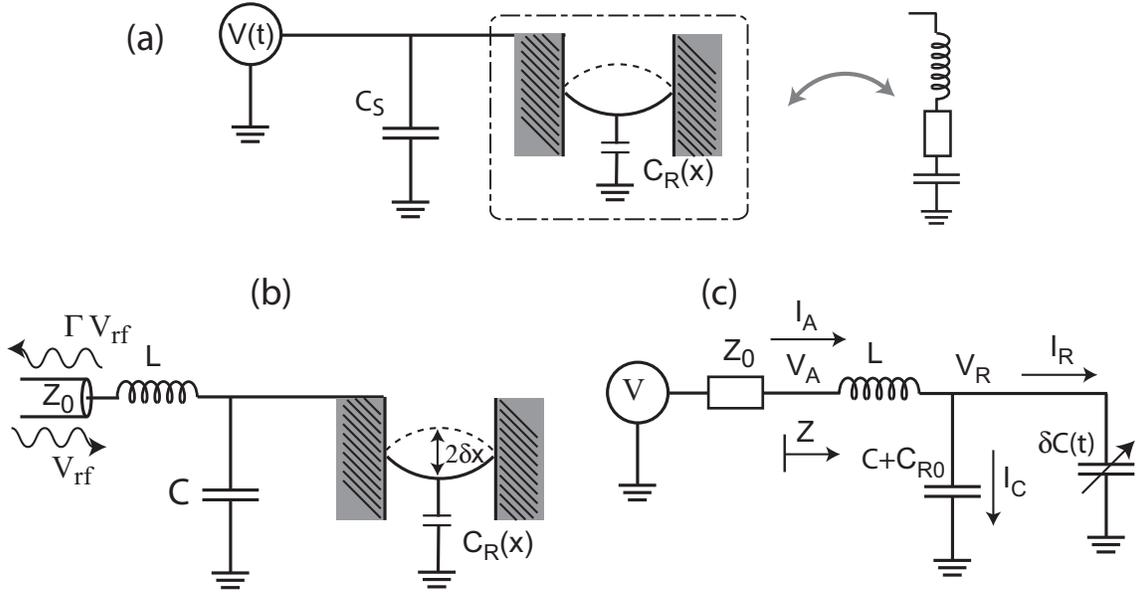}
\caption{(a) Capacitive readout of a nanomechanical resonator
(without a tank circuit) would suffer from the large parasitic
wiring capacitances $C_S$. The drive $V(t) = V_{\m{dc}} + V_0
\cos(\omega_0 t)$ is at mechanical resonance frequency $2 \pi
f_0$. Inset: electrical equivalent circuit how the mechanical
resonator looks like at the same frequency as it is driven. (b)
The signal is enhanced in our scheme by a low-$Q_{\m{LC}}$ $LC$
tank circuit which blocks the stray capacitance. The vibrations of
the beam (driven or not) at the mechanical frequency $f_0$
modulate the total capacitance of the tank circuit, and the signal
occurs as sidebands of the reflected incoming wave $V_{\m rf} \sin
\LL( \omega_{\m LC} t \RR)$ performed at the frequency $f_{\m LC}
\gg f_0$ of the tank circuit; (c) the circuit appears as one with
a time-dependent capacitance when sensed at a frequency different
from the actuation frequency $f_0$.} \label{fig:CNEMS}
\end{figure*}
%

In Fig.~\ref{fig:CNEMS} (a) we review our starting point, namely
the idea of the capacitive actuation and readout, where the NR is
driven by the voltage $V(t) = V_{\m{dc}} + V_0 \cos(\omega t)$.
The NR oscillates as a Lorentzian about the (fundamental mode)
mechanical resonance $\omega_0 = 2 \pi f_0$:
\begin{equation} \label{eq:SHOxpsol}
    \delta x (\omega, t) =  \frac{F_0 \cos(\omega t + \Theta)}{M \sqrt{\LL( \omega \omega_0/Q_M \RR)^2
    + \LL(\omega^2 - \omega_0^2\RR)^2}}
\end{equation}
where the driving ac-force is $F_0 = \LL( \frac{\partial
C_R}{\partial x}\RR) V_{\m{dc}} V_0$, $\tan \Theta = \omega_0
\omega \LL[ Q_M\LL( \omega^2 - \omega_0^2\RR) \RR]^{-1}$, $M$ is
the effective mass of the fundamental mode (about 0.73 times the
total mass of the beam), and $Q_M$ is the quality factor of the
nanomechanical mode.

Equation (\ref{eq:SHOxpsol}) gives rise to a time-varying
capacitance $C_R(t) = C_{R0} + \LL( \frac{\partial C_R}{\partial
x}\RR) \delta x$, which at the drive frequency equals the
electrical $RLC$ equivalent circuit shown in the inset of
Fig.~\ref{fig:CNEMS} (a).

On the mechanical resonance $\Theta = \pi/2$, current and voltage
are in phase and and NR looks like a resistance. Towards
increasing mechanical frequency, this effective resistance grows
rapidly, and when combined with the large wiring stray
capacitances $C_S$, makes the signal small. Truitt \emph{et
al.}~\cite{SchwabRNEMS} recently reported a progress which took
advantage of the electrical analog augmented with a tank circuit
in order to improve the impedance match to $Z_0 = 50 \, \Omega$.

In order to analyze our approach, Fig.~\ref{fig:CNEMS} (b), we
write down the time-varying capacitance when the NR is resonantly
actuated by $V(t) = V_{\m{dc}} + V_0 \cos(\omega_0 t)$:
\begin{equation} \label{eq:captime}
\begin{split}
C_R = & C_{R0} + \delta x \LL( \frac{\partial C_R}{\partial x}\RR)
\sin(\omega_0 t) = \\
= & C_{R0} + \LL( \frac{\partial C_R}{\partial
x}\RR)^2\frac{V_{\m{dc}} V_0 Q_M }{M \omega_0^2} \sin(\omega_0 t)
\end{split}
\end{equation}
Here and henceforth, we will denote by $\delta x$ the resonant
(maximum) value of the displacement in Eq.~(\ref{eq:SHOxpsol}). We
perform the measurement at a frequency $f_{\m{LC}}$ fully
different from the actuation frequency $f \sim f_0$, and as we
shall see, this has the effect of erasing the phase relationship
between the voltages and currents, and the NR therefore looks just
like a time-varying capacitance, as displayed in
Fig.~\ref{fig:CNEMS} (c). At the measurement frequency
$f_{\m{LC}}$ the $RLC$ equivalent model still holds, but the
impedance is such high that its contribution can be neglected. A
related technique was recently demonstrated by Regal \emph{et
al.}~\cite{lehnert08}, however, they used a very high-$Q$
transmission line resonator as the coupling element, and a
low-frequency NR.

For a more thorough analysis, let us write down the equations
governing the flow of information between the different
frequencies present in the problem, namely the actuation drive to
the NR at the frequency $f_0$, measurement at $f_{\m{LC}}=
\frac{1}{2\pi \sqrt{L (C + C_{R0})}}$, and the mixing products
(sidebands) $f_{\pm} \equiv f_{\m{LC}} \pm f_0$, but neglecting
higher-order mixing terms (a related analysis for the RF-SET can
be found in Ref.~\cite{LeifRFSET}).

The voltages at the various frequencies in the middle of the tank
circuit, at the point R in Fig.~\ref{fig:CNEMS} (c) are, under
these assumptions,
\begin{equation} \label{eq:VB}
\begin{split}
 V^R & = V^R_{\m{dc}} + V^R_0 \exp \LL(i \omega_0 t \RR) +
V^R_{\m{LC}} \exp \LL(i \omega_{\m{LC}} t \RR) + \\&  V^R_{+} \exp
\LL[ i (\omega_{\m{LC}} + \omega_0 ) t \RR] + V^R_{-} \exp \LL[ i
(\omega_{\m{LC}} - \omega_0 ) t \RR]
\end{split}
\end{equation}
Using Eq.~(\ref{eq:captime}) for the time-dependent capacitance
and Eq.~(\ref{eq:VB}) for the voltage across the NR, we get a
relation for the currents $I^R$ flowing through the NR:
\begin{equation} \label{eq:IR}
\begin{split}
& I^R(t)  = \frac{d}{dt} \LL[C_R(t) V^R(t) \RR] \simeq \\
& \frac{\delta x \LL( \frac{\partial C_R}{\partial x}\RR)}{2}
\Huge\{ \omega_0 V^R_{\m{dc}} \exp \LL(i \omega_0 t \RR) +
\omega_{\m{LC}} (V^R_+ - V^R_-) \exp \LL(i
\omega_{\m{LC}} t \RR) \\
&+ \omega_{+} V^R_{\m{LC}} \exp \LL(i \omega_{+} t \RR) -
\omega_{-} V^R_{\m{LC}} \exp \LL(i \omega_{-} t \RR) \Huge\} +
I_{R0}
\end{split}
\end{equation}
where $I_{R0}$ is the current through the constant part of the
capacitance $C_{R0}$. We use the Kirchoff's voltage and current
laws which allow us to solve the circuit at all the mentioned
three frequencies, when substituted by
Eqs.~(\ref{eq:VB},\ref{eq:IR}):
\begin{equation} \label{eq:kirkhoff}
\begin{split}
   V - I^A & \LL(Z_0 + i \omega L \RR) - V^R = 0 \\
  V^R = & \frac{I^A - I^R}{i \omega \LL( C + C_{R0} \RR)}
\end{split}
\end{equation}
If $\omega_0 \ll \omega_{\m{LC}}/Q_{\m{LC}}$, and in the limit of
small capacitance modulation $\delta x \LL( \frac{\partial
C_R}{\partial x}\RR) \ll C_{R0}$, we obtain from
Eqs.~(\ref{eq:kirkhoff}) the measured quantity, namely the voltage
amplitudes $V_{\pm} = Z_0 I^A_{\pm}$ of either sideband:
\begin{equation} \label{eq:Vsideband}
\begin{split}
V_{+} = & V_{-} = \frac{\delta x \LL( \frac{\partial C_R}{\partial
x}\RR) V_{\m{rf}}}{2 \omega_{\m{LC}} (C
+ C_{R0})^2 Z_0} =   \\
\simeq & \LL( \frac{C_{R0} }{C + C_{R0}} \RR)^2 \frac{Q_M
V_{\m{dc}} V_0 V_{\m{rf}}}{2 x_0^2 M \omega_{\m{LC}} Z_0
\omega_0^2}
\end{split}
\end{equation}
where for the last form, we substituted the capacitance modulation
amplitude $\delta C$ as defined in Eq.~(\ref{eq:captime}), and
then approximated it as $\frac{\partial C_R}{\partial x} \sim
\frac{C_{R0}}{x_0}$ where $x_0$ is the vacuum gap between the beam
and the gate. As one would intuitively think, the signal depends
strongly on how much stray capacitance $C$ one has \emph{within}
the resonant circuit, the part which is not cancelled by the
inductor. The signal is proportional to both actuation parameters
($V_{\m dc}$ and $V_0$) and to how much the system responds, given
by $Q_M$, as well as to the measurement strength $V_{\m rf}$.
Also, the signal can be interpreted as being proportional to the
quality factor $Q_{\m LC} \sim \LL( \omega_{\m LC} C Z_0
\RR)^{-1}$ of the tank circuit. The bandwidth, as typical of a
modulation scheme, is given by the response time of the electrical
tank circuit, as $f_{\m LC} / Q_{\m LC}$.

%
\begin{figure}[!h]
\center
\includegraphics[width=7.0cm]{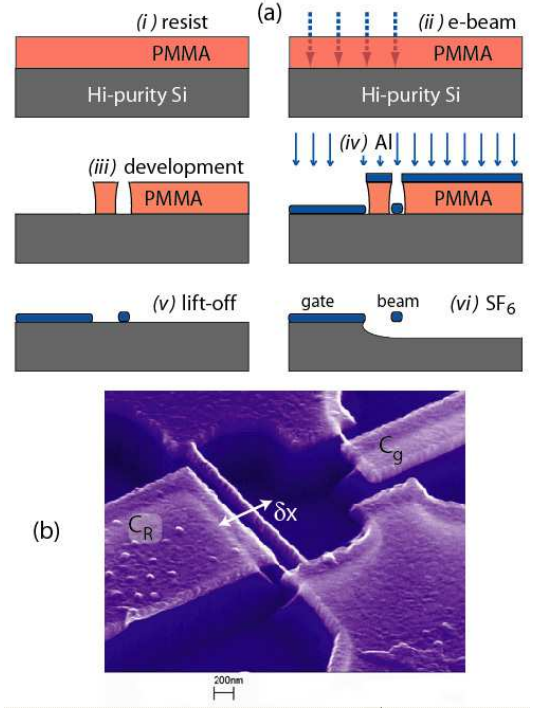}
\caption{The fabrication of the nanomechanical resonators on
high-resistivity Silicon substrate; (a) the process steps: (i)
high-purity Si substrate is coated by 300 nm of PMMA resist; (ii)
positive e-beam exposure of the final pattern; (iii) resist
development; (iv) e-beam evaporation of a thickness $h = 50$ nm or
$h = 100$ nm of Aluminum; (v) lift-off in acetone; (vi) the chip
as a whole is exposed to SF$_6$ Reactive Ion Etch (RIE) which
removes $\sim 1 \, \mu$m layer of the Si substrate. (b) SEM image
of a $l = 1.8 \, \mu$m long and $W = 160$ nm wide Al beam (similar
to samples A and B). The scale bar is 200 nm. The beam is driven
to oscillate (amplitude $\delta x$) in the plane of the film due
to capacitive drive from the large gate $C_R$. The same gate is
also used to sense the motion. The other, weakly coupled, gate
$C_g$ was not used presently.} \label{fig:CNEMSfab}
\end{figure}

For the fabrication of micron-scale suspended, metallized of fully
metallic nanomechanical beams, a multitude of ingenious methods
have been developed
\cite{cleland96Si,clelandAlNi,RoukesSiC,RoukesNanowire,parpiaDiam,sazonovaNature,pashkinNEMS}.
Our process represents the simplest end of the spectrum in terms
of the complexity of the process, and the number of steps needed.
The beam itself is metallic, and hence there is no separate
metallization needed. The fabrication goes on top of a
high-resistivity ($\sim 3$ k$\Omega$m) Silicon as shown in
Fig.~\ref{fig:CNEMSfab}, by e-beam lithography. In the end, a
properly timed SF$_6$ dry etch suspends the beam, while leaving
the clamps in both ends well hooked to the substrate.

For the measurements, the chip was wire-bonded to a surface-mount
inductor $L = 10...30$ nH in a sample holder. The tank circuit
capacitance $C \sim 0.3$ pF comes from the stray capacitances of
the bonding pad and of the inductor. The choice of the inductor
was made such that the expected mechanical frequency falls within
the electrical bandwidth $Z_0 / (2 \pi L)$. The sample was
connected to high-frequency coaxial cables in a test cryostat, and
cooled down to 4 K in a vacuum of $\sim 10^{-2}$ mBar. The
voltages at dc, at the NR drive frequency, and at the measurement
frequency were combined at room temperature using bias tees. The
signal reflected from the sample was feed to a spectrum analyzer
via a circulator and room temperature microwave amplifiers having
the noise temperature $T_N \sim 100$ K which set the noise level
in the measurement. The driven mechanical response was obtained by
scanning the drive frequency $f_g$ about the mechanical resonance
$f_0$, and recording the amplitude of the sideband voltage.

\begin{figure*}
\center
\includegraphics[width=16.0cm]{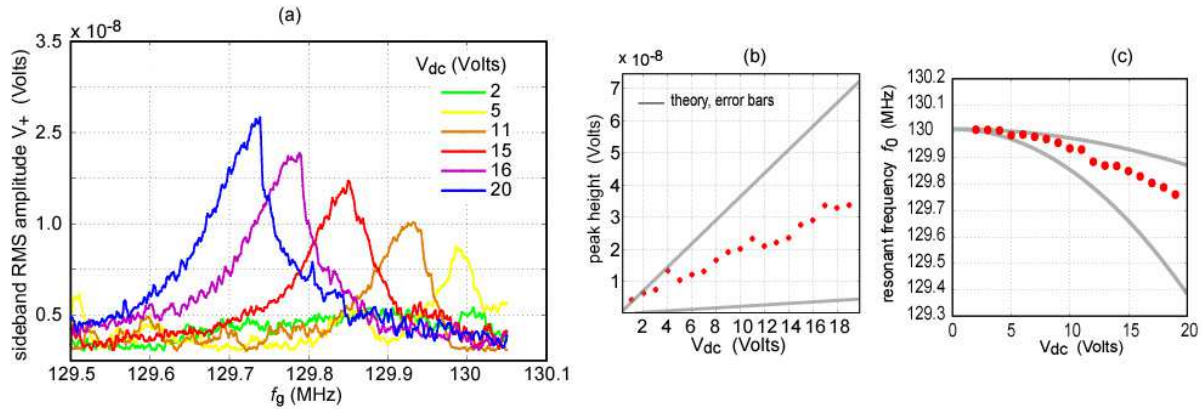}
\caption{Measurements of the response of the nanomechanical
resonator (sample D) using the electric sideband detection, $V_0 =
100$ mV and $V_{\m{LC}} = 30$ mV, at $T = 4.2$ K; (a) resonance
curves of the fundamental mode for increasing dc voltage (from
right to left); (b) Amplitude of the peak extracted from (a). The
theoretical predictions (gray lines) are scaled down by the total
microwave attenuation $\sim 10$ dB of the system; (c) center
frequency of the mechanical peak extracted from (a), compared to
theory from Eq.~(\ref{eq:wdc}); } \label{fig:CNEMSdata}
\end{figure*}

\begin{figure}[!h]
\center
\includegraphics[width=6.0cm]{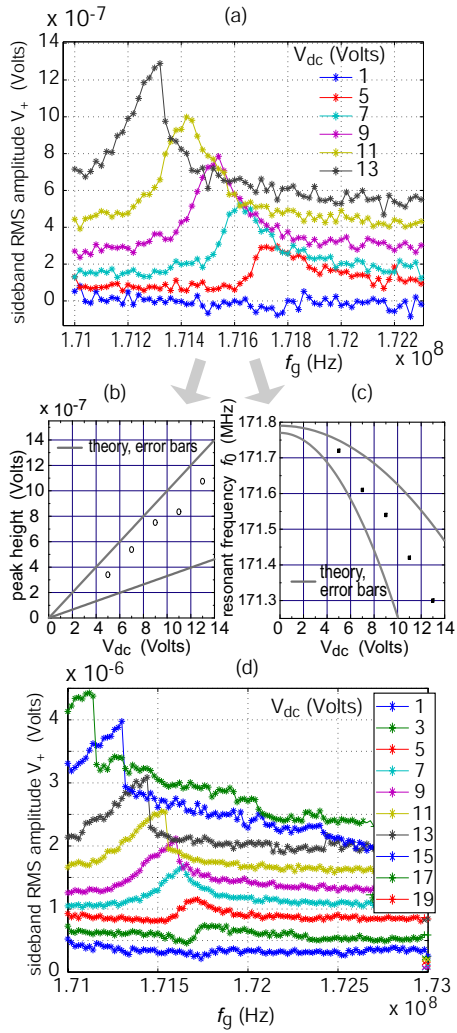}
\caption{As Fig.~\ref{fig:CNEMSdata}; measurement and analysis for
sample A; (a) $V_0 = 50$ mV and probe $V_{\m{LC}} = 140$ mV, for
increasing values of the dc voltage $V_{\m{dc}}$ from bottom to
top (displaced vertically for clarity); (b) increase of the height
of the peak versus dc voltage; (c) frequency shift versus dc
voltage; (d) at a higher total drive $V_0 = 50$ mV and $V_{\m{LC}}
= 200$ mV, the system turns nonlinear and hysteretic.}
\label{fig:CNEMSdataD}
\end{figure}

We studied a total of four samples as summarized in Table
\ref{tab:samples} for their dimensions, the used tank circuit, and
the basic characteristics $f_0$ and $Q_M$ of the resonance. In
Figs.~\ref{fig:CNEMSdata}, \ref{fig:CNEMSdataD} we show more data
for two representative samples.
\begin{table*}
\caption{\label{tab:samples} List of the measured nanomechanical
resonators. They were fabricated out of Al according to
Fig.~\ref{fig:CNEMSfab}. Sample C was baked in vacuum at 200$^o$C
for 15 min. The quality factors $Q_M$ and the fundamental mode
resonance frequencies $f_0$ were measured in a temperature of 4.2
K in a vacuum of $\sim 10^{-2}$ mBar. The estimated frequency is
given as $f_0 \simeq \frac{W}{l^2}\sqrt{\frac{E}{\rho}}$, where
$E$ is the Young modulus and $\rho$ is density. The other beam
parameters are $l = $length, $W$ = width, $h$ = thickness, $x_0$ =
vacuum gap.}
\footnotesize
\begin{ruledtabular}
\begin{tabular}{|l|c|c|c|c|c|c|c|c|c|c|c|}
sample & $l$ & $W$ & $h$ & $x_0$ & $f_0$ (est) & $f_0$ (meas)& $L$ & $C_R$ & $C$ & $f_{\m{LC}}$ & $Q_M$ \\
\hline
A & 1.8 $\mu$m & 150 nm & 100 nm & 65 nm & 230 MHz & 172 MHz & 20 nH & 47 aF & 0.39 pF & 1.81 GHz & $1.0 \times 10^3$  \\
B & 1.8 $\mu$m & 160 nm & 50 nm & 70 nm & 245 MHz & 191 MHz & 14 nH & 45 aF & 0.27 pF & 2.16 GHz & $1.0 \times 10^3$  \\
C & 2.2 $\mu$m & 200 nm & 75 nm & 100 nm & 205 MHz & 202 MHz & 10 nH & 64 aF & 0.3 pF & 3.02 GHz & $1.0 \times 10^3$  \\
D & 2.5 $\mu$m & 145 nm & 150 nm & 170 nm & 117 MHz & 130 MHz & 10 nH & 49 aF & 0.3 pF & 2.95 GHz & $2.6 \times 10^3$  \\
\end{tabular}
\end{ruledtabular}
\end{table*}
\normalsize
As the basic test of the scheme, we compared the measured sideband
voltage to that expected, Eq.~(\ref{eq:Vsideband}) according to
our model of a time-varying capacitance. We model the capacitance
$C_R$ between the beam and the nearby gate as that for two
parallel beams of length $l$, width $W$ and a vacuum gap $x_0$:
\begin{equation}\label{eq:beamCap}
    C_R(x_0) = \pi \epsilon_0 l \LL\{ \ln \LL[ \frac{x_0}{W} + 1 +
    \sqrt{\LL(\frac{x_0}{W}\RR)^2  + \frac{2x_0}{W}} \RR] \RR\}^{-1}
\end{equation}
In Figs.~\ref{fig:CNEMSdata} (b), \ref{fig:CNEMSdataD} (b) we plot
the height of the peak obtained as the drive frequency $f_g$ was
scanned through the resonance, as a function of the dc voltage. We
calculate numerically $\LL( \frac{\partial C_R}{\partial x}\RR)$
which we use in Eq.~(\ref{eq:Vsideband}) in order to obtain the
gray lines illustrating the expected linear increase of the peak
height, showing a good agreement with the data. The error bars
arise from uncertainties in the attenuation of the system at low
temperature, as well as uncertainty in the exact value of the beam
gap $x_0$ to which the capacitance is sensitive. Note, however,
that all the parameters were independently estimated.

As a further test of integrity of our model, we studied the shift
of the linear-regime resonant frequency as a function of the
applied dc voltage $V_{\m{dc}}$. Due to electrostatic softening of
the effective spring constant $k^*(V_{\m{dc}}) = k -\frac{1}{2}
V_{\m dc}^2 \frac{\partial^2 C_R}{\partial x^2}
\Large|_{V_{\m{dc}}}$, the resonance is expected to shift left
with increasing dc voltage:
\begin{equation} \label{eq:wdc}
    \omega_0 \LL( V_{\m{dc}} \RR) = \omega_0 \LL(0 \RR)
    \LL[ 1 - \frac{C_{R0} V_{\m{dc}}^2}{2 M \omega_0^2 \LL(0
    \RR)x_0^2}\RR]
\end{equation}
which holds well for the gap smaller than beam diameter. In
Figs.~\ref{fig:CNEMSdata} (c), \ref{fig:CNEMSdataD} (c) we plot
the theory prediction together with data as the gray curves which
represent Eq.~(\ref{eq:wdc}) evaluated at opposite limits of the
error bars. We notice the match is good to both the magnitude of
the frequency shift, as well as its expected quadratic behavior,
again without any fitting parameters.

The mechanical frequencies summarized in Table \ref{tab:samples}
agree within 25 \% of the prediction based on a stress-free film.
For samples A and B the measured frequency falls below the
predicted, whereas C and D show the opposite. Sample C was briefly
heat-treated (see Table \ref{tab:samples}), which could have
removed the supposedly compressive stress present in the film
after evaporation, thus causing the frequency to go up. The good
mechanical $Q$-values $Q_M > 10^3$, in agreement with previous
4-Kelvin experiments on Aluminum beams \cite{pashkinNEMS}, do not
indicate damage to the beam material was caused by the process.

While our scheme in principle offers room temperature operation,
we found out that the (small) conductivity of the Si substrate
caused spurious non-linearities at temperatures above that of
liquid nitrogen (77 K), which masked the signal. This issue could
be settled by using a more resistive substrate, such as oxidized
Silicon.


The fast sideband detection procedure which we demonstrate here
offers a sensitive, an in-principle tabletop characterization
method for the nanomechanical resonators. A major figure of merit
is the displacement sensitivity $s_x$. It is obtained from
Eq.~(\ref{eq:Vsideband}) as the value of $\delta x$ which would
correspond to a voltage spectral density equal to the noise
voltage $\sqrt{2 k_B T_N Z_0}$ which we suppose is set by the
amplifiers:
\begin{equation}\label{eq:sx}
   s_x \simeq 2 \sqrt{2k_B T_N Z_0^3}  x_0 \frac{\omega_{\m{LC}} (C
+ C_{R0})^2}{C_{R0} V_{\m{rf}}}
\end{equation}
where $T_N$ is the noise temperature of the system. Notice that
the sensitivity does not depend on the \emph{actuation} voltages
$V_{\m dc}$ and $V_0$, but evidently improves with increasing
\emph{measurement} voltage $V_{\m rf}$ under the assumption that
linearity holds. The dependence on the sample dimensions comes via
$C_{R0}$ which scales with the length $l$ as $C_{R0} \propto l$.
Hence in the "easy" limit where the stray capacitance $C$
dominates, towards small size the sensitivity degrades relatively
slowly, as $\propto l^{-2}$. For instance, we expect the method to
be usable in order to detect small vibrations for an Al doubly
clamped beam of length $l = 0.7 \, \mu$m, $f_0 \simeq 1$ GHz. Let
us use the moderate values $L = 8$ nH, $C = 0.1$ pF, $T_N = 100$
K, $V_{\m{rf}} = 3$ mV, and $V_{\m{dc}} = 10$ V, with which values
we find a sensitivity of $s_x \sim 0.5$ pm$/\sqrt{\m{Hz}}$.

The sideband readout also seems promising for basic studies of the
NR at low temperatures, since the NR and its surroundings stay
superconducting. The dispersive detection where only the phase of
the reflected signal varies, as well as the all-electric actuation
we apply, cause a vanishing on-chip dissipation.

Interestingly, the sideband measurement might also offer a path
towards the quantum limit of mechanical motion. Through
minimization of the tank circuit capacitance $C$ in
Fig.~\ref{fig:CNEMS} (b) by improving the $LC$ tank circuit from
our first realization made of surface mount components, one could
approach the sensitivity needed for observing the mechanical
zero-point vibrations. The lowest $C$ would be achieved by
fabricating the NR coupled to an on-chip spiral coil, for example
$L = 17$ nH, $C \sim 25$ fF. Let us consider an Aluminum beam
having a length $l = 1.5 \, \mu$m and frequency $f_0 = 240$ MHz.
Let us also take $f_{\m LC} = 5$ GHz and $T_N = 4$ K. If probed
even with a decent $V_{\m{rf}} = 3$ mV, and $V_{\m{dc}} = 10$ V,
Eq.~(\ref{eq:sx}) yields a displacement sensitivity $s_x \sim 45$
fm$/\sqrt{\m{Hz}}$, which comes interestingly close to the
zero-point aplitude $x_{\m{zp}} \sim 25$ fm whose energy will be
spread over a bandwidth determined by $Q_M$. The back-action
disturbance due to the probing remains small, since $f_{\m{LC}}
\gg f_0$. The back-action is estimated from the tail of the
Lorentzian response of Eq.~(\ref{eq:SHOxpsol}), where the
amplitude at high frequencies decays as $\LL( f_{\m{LC}}
\RR)^{-3}$. With the above parameter values, we find that the
measurement would excite less than one quantum of energy into the
NR.

An interesting materials platform for the sideband readout, among
all conductive materials, is graphene \cite{ParpiaGraph}. The
signal improves due to a large capacitance of the sheet and due to
its low mass, whereas the latter also substantially increases the
zero-point amplitude. Due to the sensitivity of the novel method,
interesting mass, acceleration or other sensor applications might
also arise.

\begin{acknowledgments}
We wish to acknowledge Jukka Pekola, Sami Franssila and Antti O.
Niskanen for useful discussions. This work was supported by the
Academy of Finland, and by EU contract FP6-021285.
\end{acknowledgments}

\bibliography{c:/mika/latex/mikabib}

\begin{thebibliography}{24}
\expandafter\ifx\csname natexlab\endcsname\relax\def\natexlab#1{#1}\fi
\expandafter\ifx\csname bibnamefont\endcsname\relax
  \def\bibnamefont#1{#1}\fi
\expandafter\ifx\csname bibfnamefont\endcsname\relax
  \def\bibfnamefont#1{#1}\fi
\expandafter\ifx\csname citenamefont\endcsname\relax
  \def\citenamefont#1{#1}\fi
\expandafter\ifx\csname url\endcsname\relax
  \def\url#1{\texttt{#1}}\fi
\expandafter\ifx\csname urlprefix\endcsname\relax\def\urlprefix{URL }\fi
\providecommand{\bibinfo}[2]{#2}
\providecommand{\eprint}[2][]{\url{#2}}

\bibitem[{\citenamefont{Cleland}(2003)}]{clelandbook}
\bibinfo{author}{\bibfnamefont{A.}~\bibnamefont{Cleland}},
  \emph{\bibinfo{title}{Foundations of Nanomechanics}}
  (\bibinfo{publisher}{Springer}, \bibinfo{address}{New York},
  \bibinfo{year}{2003}).

\bibitem[{\citenamefont{Ekinci and Roukes}(2005)}]{RoukesReview}
\bibinfo{author}{\bibfnamefont{K.~L.} \bibnamefont{Ekinci}} \bibnamefont{and}
  \bibinfo{author}{\bibfnamefont{M.~L.} \bibnamefont{Roukes}},
  \bibinfo{journal}{Rev. Sci. Instrum.} \textbf{\bibinfo{volume}{76}},
  \bibinfo{pages}{061101} (\bibinfo{year}{2005}).

\bibitem[{\citenamefont{Jensen et~al.}(2008)\citenamefont{Jensen, Kim, and
  Zettl}}]{NRsensor}
\bibinfo{author}{\bibfnamefont{K.}~\bibnamefont{Jensen}},
  \bibinfo{author}{\bibfnamefont{K.}~\bibnamefont{Kim}}, \bibnamefont{and}
  \bibinfo{author}{\bibfnamefont{A.}~\bibnamefont{Zettl}},
  \bibinfo{journal}{Nature Nanotech.} \textbf{\bibinfo{volume}{3}},
  \bibinfo{pages}{533} (\bibinfo{year}{2008}).

\bibitem[{\citenamefont{Irish and Schwab}(2003)}]{Schwab03}
\bibinfo{author}{\bibfnamefont{E.~K.} \bibnamefont{Irish}} \bibnamefont{and}
  \bibinfo{author}{\bibfnamefont{K.}~\bibnamefont{Schwab}},
  \bibinfo{journal}{Phys. Rev. B} \textbf{\bibinfo{volume}{68}},
  \bibinfo{pages}{155311} (\bibinfo{year}{2003}).

\bibitem[{\citenamefont{Cleland and Geller}(2004)}]{cleland04}
\bibinfo{author}{\bibfnamefont{A.~N.} \bibnamefont{Cleland}} \bibnamefont{and}
  \bibinfo{author}{\bibfnamefont{M.~R.} \bibnamefont{Geller}},
  \bibinfo{journal}{Phys. Rev. Lett.} \textbf{\bibinfo{volume}{93}},
  \bibinfo{pages}{070501} (\bibinfo{year}{2004}).

\bibitem[{\citenamefont{Greywall et~al.}(1994)\citenamefont{Greywall, Yurke,
  Busch, Pargellis, and Willett}}]{YurkeMagn94}
\bibinfo{author}{\bibfnamefont{D.~S.} \bibnamefont{Greywall}},
  \bibinfo{author}{\bibfnamefont{B.}~\bibnamefont{Yurke}},
  \bibinfo{author}{\bibfnamefont{P.~A.} \bibnamefont{Busch}},
  \bibinfo{author}{\bibfnamefont{A.~N.} \bibnamefont{Pargellis}},
  \bibnamefont{and} \bibinfo{author}{\bibfnamefont{R.~L.}
  \bibnamefont{Willett}}, \bibinfo{journal}{Phys. Rev. Lett.}
  \textbf{\bibinfo{volume}{72}}, \bibinfo{pages}{2992} (\bibinfo{year}{1994}).

\bibitem[{\citenamefont{Cleland and Roukes}(1996)}]{cleland96Si}
\bibinfo{author}{\bibfnamefont{A.~N.} \bibnamefont{Cleland}} \bibnamefont{and}
  \bibinfo{author}{\bibfnamefont{M.~L.} \bibnamefont{Roukes}},
  \bibinfo{journal}{Appl. Phys. Lett.} \textbf{\bibinfo{volume}{69}},
  \bibinfo{pages}{2653} (\bibinfo{year}{1996}).

\bibitem[{\citenamefont{Ekinci et~al.}(2002)\citenamefont{Ekinci, Yang, Huang,
  and Roukes}}]{RoukesBalance}
\bibinfo{author}{\bibfnamefont{K.~L.} \bibnamefont{Ekinci}},
  \bibinfo{author}{\bibfnamefont{Y.~T.} \bibnamefont{Yang}},
  \bibinfo{author}{\bibfnamefont{X.~M.~H.} \bibnamefont{Huang}},
  \bibnamefont{and} \bibinfo{author}{\bibfnamefont{M.~L.}
  \bibnamefont{Roukes}}, \bibinfo{journal}{Appl. Phys. Lett.}
  \textbf{\bibinfo{volume}{81}}, \bibinfo{pages}{2253} (\bibinfo{year}{2002}).

\bibitem[{\citenamefont{Li et~al.}(2008)\citenamefont{Li, Pashkin, Astafiev,
  Nakamura, Tsai, and Im}}]{pashkinNEMS}
\bibinfo{author}{\bibfnamefont{T.~F.} \bibnamefont{Li}},
  \bibinfo{author}{\bibfnamefont{Y.~A.} \bibnamefont{Pashkin}},
  \bibinfo{author}{\bibfnamefont{O.}~\bibnamefont{Astafiev}},
  \bibinfo{author}{\bibfnamefont{Y.}~\bibnamefont{Nakamura}},
  \bibinfo{author}{\bibfnamefont{J.~S.} \bibnamefont{Tsai}}, \bibnamefont{and}
  \bibinfo{author}{\bibfnamefont{H.}~\bibnamefont{Im}}, \bibinfo{journal}{Appl.
  Phys. Lett.} \textbf{\bibinfo{volume}{92}}, \bibinfo{pages}{043112}
  (\bibinfo{year}{2008}).

\bibitem[{\citenamefont{LaHaye et~al.}(2004)\citenamefont{LaHaye, Buu,
  Camarota, and Schwab}}]{lahaye04}
\bibinfo{author}{\bibfnamefont{M.~D.} \bibnamefont{LaHaye}},
  \bibinfo{author}{\bibfnamefont{O.}~\bibnamefont{Buu}},
  \bibinfo{author}{\bibfnamefont{B.}~\bibnamefont{Camarota}}, \bibnamefont{and}
  \bibinfo{author}{\bibfnamefont{K.~C.} \bibnamefont{Schwab}},
  \bibinfo{journal}{Science} \textbf{\bibinfo{volume}{304}},
  \bibinfo{pages}{74} (\bibinfo{year}{2004}).

\bibitem[{\citenamefont{Naik et~al.}(2006)\citenamefont{Naik, Buu, LaHaye,
  Armour, Clerk, Blencowe, and Schwab}}]{Naik06}
\bibinfo{author}{\bibfnamefont{A.}~\bibnamefont{Naik}},
  \bibinfo{author}{\bibfnamefont{O.}~\bibnamefont{Buu}},
  \bibinfo{author}{\bibfnamefont{M.~D.} \bibnamefont{LaHaye}},
  \bibinfo{author}{\bibfnamefont{A.~D.} \bibnamefont{Armour}},
  \bibinfo{author}{\bibfnamefont{A.~A.} \bibnamefont{Clerk}},
  \bibinfo{author}{\bibfnamefont{M.~P.} \bibnamefont{Blencowe}},
  \bibnamefont{and} \bibinfo{author}{\bibfnamefont{K.~C.}
  \bibnamefont{Schwab}}, \bibinfo{journal}{Nature}
  \textbf{\bibinfo{volume}{443}}, \bibinfo{pages}{193} (\bibinfo{year}{2006}).

\bibitem[{\citenamefont{Regal et~al.}(2008)\citenamefont{Regal, Teufel, and
  Lehnert}}]{lehnert08}
\bibinfo{author}{\bibfnamefont{J.~D.} \bibnamefont{Regal}},
  \bibinfo{author}{\bibfnamefont{C.~A.} \bibnamefont{Teufel}},
  \bibnamefont{and} \bibinfo{author}{\bibfnamefont{K.~W.}
  \bibnamefont{Lehnert}}, \bibinfo{journal}{Nature Physics}
  \textbf{\bibinfo{volume}{4}}, \bibinfo{pages}{555} (\bibinfo{year}{2008}).

\bibitem[{\citenamefont{Huang et~al.}(2003)\citenamefont{Huang, Zorman,
  Mehregany, and Roukes}}]{RoukesGHz}
\bibinfo{author}{\bibfnamefont{X.~M.~H.} \bibnamefont{Huang}},
  \bibinfo{author}{\bibfnamefont{C.~A.} \bibnamefont{Zorman}},
  \bibinfo{author}{\bibfnamefont{M.}~\bibnamefont{Mehregany}},
  \bibnamefont{and} \bibinfo{author}{\bibfnamefont{M.~L.}
  \bibnamefont{Roukes}}, \bibinfo{journal}{Nature}
  \textbf{\bibinfo{volume}{421}}, \bibinfo{pages}{496} (\bibinfo{year}{2003}).

\bibitem[{\citenamefont{Truitt et~al.}(2007)\citenamefont{Truitt, Hertzberg,
  Huang, Ekinci, and Schwab}}]{SchwabRNEMS}
\bibinfo{author}{\bibfnamefont{P.~A.} \bibnamefont{Truitt}},
  \bibinfo{author}{\bibfnamefont{J.~B.} \bibnamefont{Hertzberg}},
  \bibinfo{author}{\bibfnamefont{C.~C.} \bibnamefont{Huang}},
  \bibinfo{author}{\bibfnamefont{K.~L.} \bibnamefont{Ekinci}},
  \bibnamefont{and} \bibinfo{author}{\bibfnamefont{K.~C.}
  \bibnamefont{Schwab}}, \bibinfo{journal}{Nano Lett.}
  \textbf{\bibinfo{volume}{7}}, \bibinfo{pages}{120} (\bibinfo{year}{2007}).

\bibitem[{\citenamefont{Schoelkopf et~al.}(1998)\citenamefont{Schoelkopf,
  Wahlgren, Kozhevnikov, Delsing, and Prober}}]{rfset}
\bibinfo{author}{\bibfnamefont{R.}~\bibnamefont{Schoelkopf}},
  \bibinfo{author}{\bibfnamefont{P.}~\bibnamefont{Wahlgren}},
  \bibinfo{author}{\bibfnamefont{A.}~\bibnamefont{Kozhevnikov}},
  \bibinfo{author}{\bibfnamefont{P.}~\bibnamefont{Delsing}}, \bibnamefont{and}
  \bibinfo{author}{\bibfnamefont{D.}~\bibnamefont{Prober}},
  \bibinfo{journal}{Science} \textbf{\bibinfo{volume}{280}},
  \bibinfo{pages}{1238} (\bibinfo{year}{1998}).

\bibitem[{\citenamefont{Roschier et~al.}(2005)\citenamefont{Roschier,
  Sillanp\"{a}\"{a}, and Hakonen}}]{CSET}
\bibinfo{author}{\bibfnamefont{L.}~\bibnamefont{Roschier}},
  \bibinfo{author}{\bibfnamefont{M.}~\bibnamefont{Sillanp\"{a}\"{a}}},
  \bibnamefont{and} \bibinfo{author}{\bibfnamefont{P.}~\bibnamefont{Hakonen}},
  \bibinfo{journal}{Phys. Rev. B} \textbf{\bibinfo{volume}{71}},
  \bibinfo{pages}{024530} (\bibinfo{year}{2005}).

\bibitem[{\citenamefont{Sillanp\"{a}\"{a}
  et~al.}(2004)\citenamefont{Sillanp\"{a}\"{a}, Roschier, and
  Hakonen}}]{sillanpaa04}
\bibinfo{author}{\bibfnamefont{M.}~\bibnamefont{Sillanp\"{a}\"{a}}},
  \bibinfo{author}{\bibfnamefont{L.}~\bibnamefont{Roschier}}, \bibnamefont{and}
  \bibinfo{author}{\bibfnamefont{P.}~\bibnamefont{Hakonen}},
  \bibinfo{journal}{Phys. Rev. Lett.} \textbf{\bibinfo{volume}{93}},
  \bibinfo{pages}{066805} (\bibinfo{year}{2004}).

\bibitem[{\citenamefont{Roschier et~al.}(2004)\citenamefont{Roschier,
  Sillanp\"{a}\"{a}, Wang, Ahlskog, Iijima, and Hakonen}}]{LeifRFSET}
\bibinfo{author}{\bibfnamefont{L.}~\bibnamefont{Roschier}},
  \bibinfo{author}{\bibfnamefont{M.}~\bibnamefont{Sillanp\"{a}\"{a}}},
  \bibinfo{author}{\bibfnamefont{T.}~\bibnamefont{Wang}},
  \bibinfo{author}{\bibfnamefont{M.}~\bibnamefont{Ahlskog}},
  \bibinfo{author}{\bibfnamefont{S.}~\bibnamefont{Iijima}}, \bibnamefont{and}
  \bibinfo{author}{\bibfnamefont{P.}~\bibnamefont{Hakonen}},
  \bibinfo{journal}{Journal of Low Temperature Physics}
  \textbf{\bibinfo{volume}{136}}, \bibinfo{pages}{465} (\bibinfo{year}{2004}).

\bibitem[{\citenamefont{Cleland et~al.}(2001)\citenamefont{Cleland, Pophristic,
  and Ferguson}}]{clelandAlNi}
\bibinfo{author}{\bibfnamefont{A.~N.} \bibnamefont{Cleland}},
  \bibinfo{author}{\bibfnamefont{M.}~\bibnamefont{Pophristic}},
  \bibnamefont{and} \bibinfo{author}{\bibfnamefont{I.}~\bibnamefont{Ferguson}},
  \bibinfo{journal}{Appl. Phys. Lett.} \textbf{\bibinfo{volume}{79}},
  \bibinfo{pages}{2070} (\bibinfo{year}{2001}).

\bibitem[{\citenamefont{Yang et~al.}(2001)\citenamefont{Yang, Ekinci, Huang,
  Schiavone, Roukes, Zorman, and Mehregany}}]{RoukesSiC}
\bibinfo{author}{\bibfnamefont{Y.~T.} \bibnamefont{Yang}},
  \bibinfo{author}{\bibfnamefont{K.~L.} \bibnamefont{Ekinci}},
  \bibinfo{author}{\bibfnamefont{X.~M.~H.} \bibnamefont{Huang}},
  \bibinfo{author}{\bibfnamefont{L.~M.} \bibnamefont{Schiavone}},
  \bibinfo{author}{\bibfnamefont{M.~L.} \bibnamefont{Roukes}},
  \bibinfo{author}{\bibfnamefont{C.~A.} \bibnamefont{Zorman}},
  \bibnamefont{and}
  \bibinfo{author}{\bibfnamefont{M.}~\bibnamefont{Mehregany}},
  \bibinfo{journal}{Appl. Phys. Lett.} \textbf{\bibinfo{volume}{78}},
  \bibinfo{pages}{162} (\bibinfo{year}{2001}).

\bibitem[{\citenamefont{Husain et~al.}(2003)\citenamefont{Husain, Hone, Postma,
  Huang, Drake, Barbic, Scherer, and Roukes}}]{RoukesNanowire}
\bibinfo{author}{\bibfnamefont{A.}~\bibnamefont{Husain}},
  \bibinfo{author}{\bibfnamefont{J.}~\bibnamefont{Hone}},
  \bibinfo{author}{\bibfnamefont{H.~W.~C.} \bibnamefont{Postma}},
  \bibinfo{author}{\bibfnamefont{X.~M.~H.} \bibnamefont{Huang}},
  \bibinfo{author}{\bibfnamefont{T.}~\bibnamefont{Drake}},
  \bibinfo{author}{\bibfnamefont{M.}~\bibnamefont{Barbic}},
  \bibinfo{author}{\bibfnamefont{A.}~\bibnamefont{Scherer}}, \bibnamefont{and}
  \bibinfo{author}{\bibfnamefont{M.~L.} \bibnamefont{Roukes}},
  \bibinfo{journal}{Appl. Phys. Lett.} \textbf{\bibinfo{volume}{83}},
  \bibinfo{pages}{1240} (\bibinfo{year}{2003}).

\bibitem[{\citenamefont{Sekaric et~al.}(2002)\citenamefont{Sekaric, Parpia,
  Craighead, Feygelson, Houston, and Butler}}]{parpiaDiam}
\bibinfo{author}{\bibfnamefont{L.}~\bibnamefont{Sekaric}},
  \bibinfo{author}{\bibfnamefont{J.~M.} \bibnamefont{Parpia}},
  \bibinfo{author}{\bibfnamefont{H.~G.} \bibnamefont{Craighead}},
  \bibinfo{author}{\bibfnamefont{T.}~\bibnamefont{Feygelson}},
  \bibinfo{author}{\bibfnamefont{B.~H.} \bibnamefont{Houston}},
  \bibnamefont{and} \bibinfo{author}{\bibfnamefont{J.~E.}
  \bibnamefont{Butler}}, \bibinfo{journal}{Appl. Phys. Lett.}
  \textbf{\bibinfo{volume}{81}}, \bibinfo{pages}{4455} (\bibinfo{year}{2002}).

\bibitem[{\citenamefont{Sazonova et~al.}(2004)\citenamefont{Sazonova, Yaish,
  H.~\"Ust\"unel, and McEuen}}]{sazonovaNature}
\bibinfo{author}{\bibfnamefont{V.}~\bibnamefont{Sazonova}},
  \bibinfo{author}{\bibfnamefont{Y.}~\bibnamefont{Yaish}},
  \bibinfo{author}{\bibfnamefont{T.~A.} \bibnamefont{H.~\"Ust\"unel},
  \bibfnamefont{D.~Roundy}}, \bibnamefont{and}
  \bibinfo{author}{\bibfnamefont{P.}~\bibnamefont{McEuen}},
  \bibinfo{journal}{Nature} \textbf{\bibinfo{volume}{431}},
  \bibinfo{pages}{284} (\bibinfo{year}{2004}).

\bibitem[{\citenamefont{Bunch et~al.}(2007)\citenamefont{Bunch, van~der Zande,
  Verbridge, Frank, Tanenbaum, Parpia, Craighead, and McEuen}}]{ParpiaGraph}
\bibinfo{author}{\bibfnamefont{J.~S.} \bibnamefont{Bunch}},
  \bibinfo{author}{\bibfnamefont{A.~M.} \bibnamefont{van~der Zande}},
  \bibinfo{author}{\bibfnamefont{S.~S.} \bibnamefont{Verbridge}},
  \bibinfo{author}{\bibfnamefont{I.~W.} \bibnamefont{Frank}},
  \bibinfo{author}{\bibfnamefont{D.~M.} \bibnamefont{Tanenbaum}},
  \bibinfo{author}{\bibfnamefont{J.~M.} \bibnamefont{Parpia}},
  \bibinfo{author}{\bibfnamefont{H.~G.} \bibnamefont{Craighead}},
  \bibnamefont{and} \bibinfo{author}{\bibfnamefont{P.~L.}
  \bibnamefont{McEuen}}, \bibinfo{journal}{Science}
  \textbf{\bibinfo{volume}{315}}, \bibinfo{pages}{490} (\bibinfo{year}{2007}).

\end{thebibliography}

\end{document}